\documentclass[10pt, aps, prl,twocolumn,superscriptaddress]{revtex4-1}

\usepackage{graphicx}
\usepackage{hyperref}
\usepackage[all]{hypcap}

\hypersetup{colorlinks=true,linkcolor=blue,citecolor=blue,urlcolor=blue} 

\newcommand{\mytilde}{\raise.17ex\hbox{$\scriptstyle\mathtt{\sim}$}}
\newcommand{\USASK}{Department of Physics \& Engineering Physics, University of Saskatchewan, Saskatoon, Canada, S7N 5E2}
\newcommand{\CLS}{Canadian Light Source, University of Saskatchewan, Saskatoon, Saskatchewan S7N 0X4, Canada}
\newcommand{\QMI}{Quantum Matter Institute, University of British Columbia, Vancouver, Canada, V6T 1Z1}
\newcommand{\SUZHOU}{College of Nano Science and Technology, Soochow University, 199 Ren-Ai Rd., Suzhou Industrial Park, Suzhou, Jiangsu, 215123, China}
\newcommand{\SHANXI}{School of Chemistry \& Materials Science, Shanxi Normal University, Linfen 041004, People's Republic of China}
\newcommand{\IMP}{M. N. Mikheev Institute of Metal Physics of the Ural Branch of Russian Academy of Sciences, 620990 Yekaterinburg, Russia}
\newcommand{\UFU}{Ural Federal University, 19 Mira Str., 620002 Yekaterinburg, Russia}
\begin{document}

\title{Adjacent Fe-Vacancy Interactions as the Origin of Room Temperature Ferromagnetism in (In$_{1-x}$Fe$_{x}$)$_{2}$O$_{3}$}

\author{R. J. Green}
\email[]{robert.green@usask.ca}
\altaffiliation[Current address: ]{\QMI}
\affiliation{\USASK}

\author{T. Z. Regier}
\affiliation{\CLS}

\author{B. Leedahl}
\affiliation{\USASK}

\author{J. A. McLeod}
\altaffiliation[Current address: ]{\SUZHOU}
\affiliation{\USASK}

\author{X. H. Xu}
\affiliation{\SHANXI}

\author{G. S. Chang}
\affiliation{\USASK}

\author{E. Z. Kurmaev}
\affiliation{\IMP}
\affiliation{\UFU}

\author{A. Moewes}
\affiliation{\USASK}

\date{\today}

\begin{abstract}
Dilute magnetic semiconductors (DMSs) show great promise for applications in spin-based electronics, but in most cases continue to elude explanations of their magnetic behavior. Here, we combine quantitative x-ray spectroscopy and Anderson impurity model calculations to study ferromagnetic Fe-substituted In$_{2}$O$_{3}$ films, and we identify a subset of Fe atoms adjacent to oxygen vacancies in the crystal lattice which are responsible for the observed room temperature ferromagnetism. Using resonant inelastic x-ray scattering, we map out the near gap electronic structure and provide further support for this conclusion. Serving as a concrete verification of recent theoretical results and indirect experimental evidence, these results solidify the role of impurity-vacancy coupling in oxide-based DMSs.
\end{abstract}


\maketitle

The enigmatic nature of dilute magnetic semiconductors (DMSs) has provided an intriguing and popular topic in materials science and condensed matter physics research for well over a decade \cite{ogale2010}. Research interest in DMSs is fueled by the desire to find suitable ferromagnetic semiconductors for spintronics applications which, if realized on a large scale, would revolutionize computing capabilities \cite{Wolf2001}. Experimental reports of the desirable room temperature ferromagnetism (RTFM) obtained by substituting transition metals into semiconducting and insulating oxides are now prevalent (see, for example, Refs. \onlinecite{Ogale_PRL_2003, Ciatto_PRL_2011, Matsumoto2001, Sharma2003, Coey_2005_Nat_Mat, Ueda_APL_2001}).  What is not prevalent, however, is a universally accepted description of the mechanism which mediates the often unexpected ferromagnetic behavior. Such an understanding is necessary to develop and optimize effective spintronic devices using these materials.

Much of the original interest in DMSs centered on $p$-type materials, motivated by the discovery of magnetism at low temperatures in Ga$_{1-x}$Mn$_{x}$As \cite{Ohno_ScienceRev_1998,Ohno_GaMnAs_APL_1996}. Further interest was stimulated by a theoretical prediction of attainable RTFM in Mn-substituted, $p$-type DMS materials \cite{Dietl2000}. Recent studies of Ga$_{1-x}$Mn$_{x}$As have yielded some important developments regarding the magnetic mechanisms and electronic structure \cite{Fujii_PRL_2011, Berciu_PRL_2009, Kobayashi_GaMnAs_PRL_2014, Dobrowolska_NatMat_2012, GrayFadley_NatMat_2012}. For this material, the FM seems to be intricately linked to the tightly bound \cite{Kobayashi_GaMnAs_PRL_2014} holes in the Mn-induced impurity band which overlaps the Fermi level \cite{Dobrowolska_NatMat_2012,GrayFadley_NatMat_2012}. Unfortunately, however, to date all variations of Ga$_{1-x}$Mn$_{x}$As only exhibit FM at low temperatures. Thus, while much can be learned from studying the physics of this material, it is not an ideal candidate for spintronics applications.

Recently, the $n$-type (In$_{1-x}$Fe$_x$)$_2$O$_3$, along with other oxide-based materials, has generated significant interest due to numerous independent reports of RTFM \cite{Xu_Gehring_PRB_2011, Xu_Gehring_APL_2009, Xu_Gehring_JAP_2011, Singhal_In2O3Fe_2010, Hu_In2O3Fe_APL_2007, Chu_APL_2007_EXP_In2O3Fe, Li_In2O3Fe_APL_2005,Xing_In2O3Fe_vacAnneal_2009,An_In2O3Fe_ASS_2013,An_In2O3Fe_JPCC_2015}. In search of an explanation of the magnetism, some of these studies have provided indirect evidence that oxygen site vacancies ($V_O$) are important for the ferromagnetism. In some cases, annealing cycles between oxygen and vacuum environments have been reported to respectively destroy and restore the room temperature ferromagnetic ordering \cite{Singhal_In2O3Fe_2010, Li_In2O3Fe_APL_2005, Xing_In2O3Fe_vacAnneal_2009}. Oxygen vacancies have been considered as explanations for ferromagnetism in other DMSs as well. For example, indirect experimental evidence recently showed that interactions between Co atoms and $V_O$ in Co-substituted ZnO were integral to the observed ferromagnetism \cite{Ciatto_PRL_2011}, while defect-free samples have been found to show only paramagnetic behavior \citep{2011_NJP_Ney}.

If $V_O$ centers are important for RTFM in DMS oxides, then cubic In$_2$O$_3$ serves as a prime host for synthesis, as $V_O$ are typical native defects and participate actively in the electronic structure, leading to the naturally observed $n$-type conduction. However, like DMSs, the properties of In$_2$O$_3$ itself have been heavily debated in recent years: only recently have studies clarified the nature and magnitude of the electronic band gap \cite{Walsh_PRL_2008, King_PRB_2009}, and there has been debate regarding whether $V_O$ are deep or shallow donors and their exact role in the defect-based conductivity \cite{Zunger_In2O3_PRL_2007, Agoston_In2O3_PRL_2009, Zunger_In2O3Comment_PRL_2011, Agoston_Reply_PRL_2011}.

\begin{figure}
\includegraphics[width=3.375in]{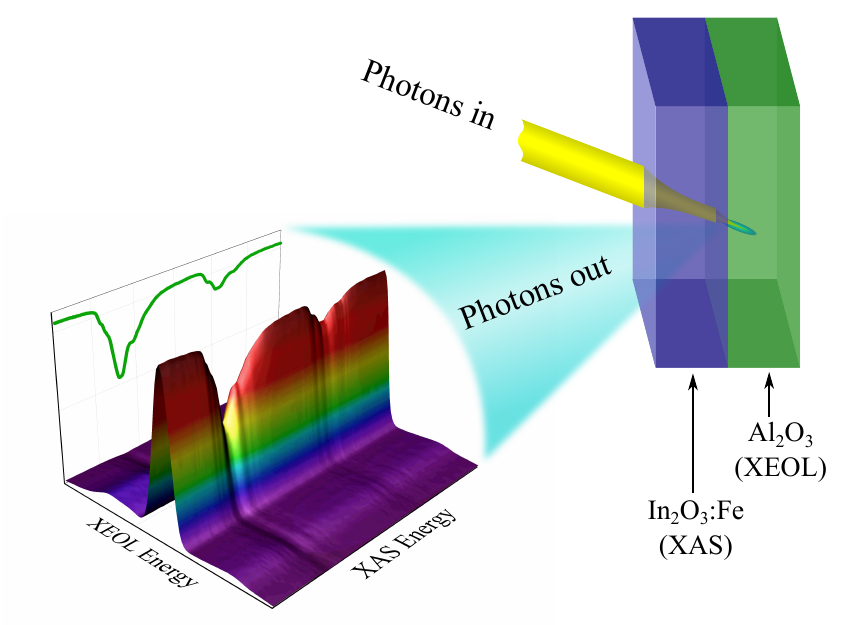}%
\caption{(Color online) Experimental setup for XAS measurements. The Al$_2$O$_3$ substrate XEOL intensity is used as a transmission detector for the (In$_{1-x}$Fe$_x$)$_2$O$_3$ film XAS.}
\label{Fig:EXP}
\end{figure}

\begin{figure*}
\includegraphics[width=7.0in]{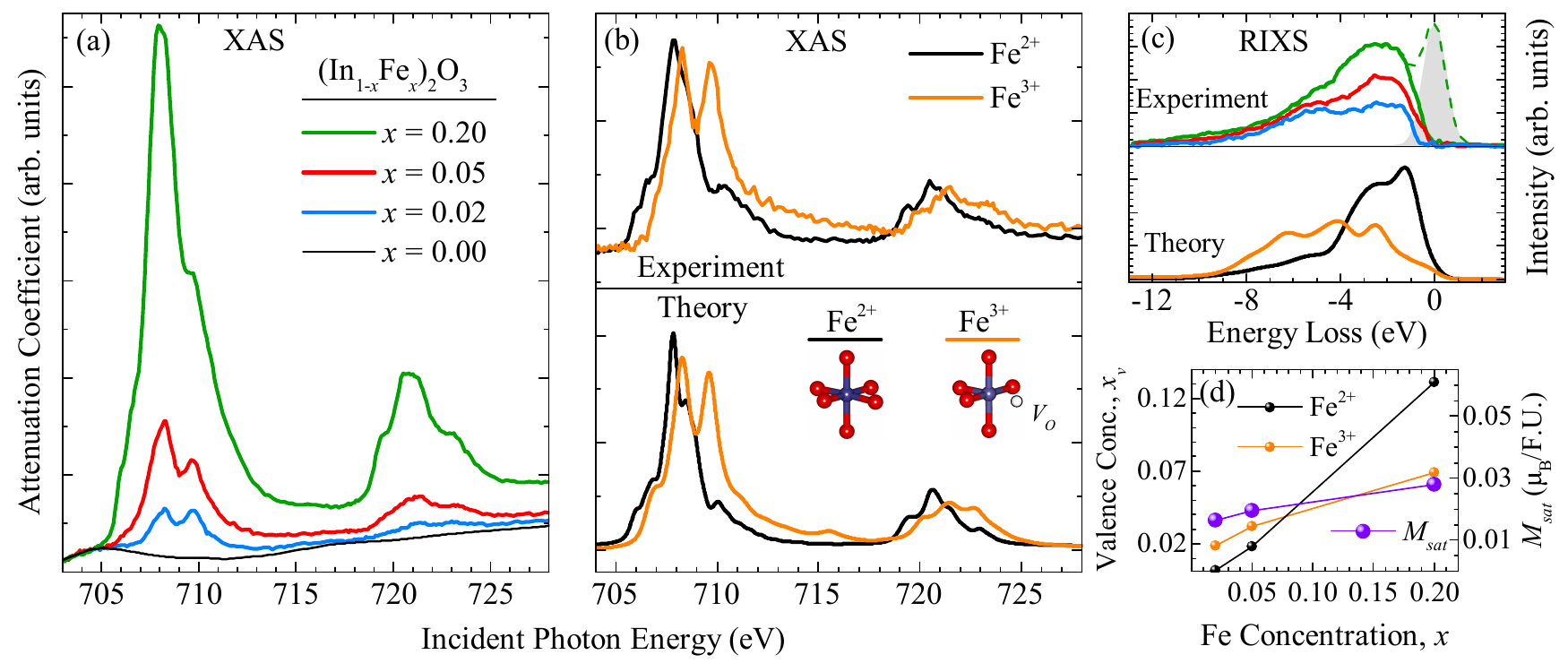}
\caption{(Color online) Fe $L_{2,3}$-edge spectroscopy results. (a) The raw XAS spectra obtained from the substrate XEOL yield as depicted in Fig \ref{Fig:EXP}. (b) The extracted component XAS spectra (upper panel) with the corresponding calculated spectra (lower panel). (c) Experimental (upper) and calculated (lower) $L_3$ RIXS spectra. The elastic contributions have been removed for all samples as demonstrated by the dashed line (raw) and grey area (elastic component removed) for one sample. The calculations use the same models as in (b). (d) The saturation magnetizations ($M_{sat}$) measured for each sample are correlated with the fractional concentration of Fe$^{3+}$-V$_O$ centers determined from XAS.}
\label{In2O3Fe_Iron}
\end{figure*}

In this work we find, using quantitative Fe $L_{2,3}$ x-ray absorption spectroscopy (XAS) and resonant inelastic x-ray scattering (RIXS), that the ferromagnetism of (In$_{1-x}$Fe$_x$)$_2$O$_3$ is linked with substitutional Fe atoms adjacent to $V_O$ centers, as predicted by recent \emph{ab initio} calculations \cite{Xu_Gehring_PRB_2011}. We find that the the concentration of these Fe-$V_{O}$ centers correlates directly to the magnitude of the saturation magnetization measured for a series of high quality thin film samples. We use oxygen $K$-edge XAS and RIXS to reveal the interaction between the Fe $3d$ electronic states and the host In$_2$O$_3$ valence and conduction bands. We find that it is a strong local interaction, facilitated by the Fe-$V_{O}$ proximity, that mediates the ferromagnetism through resonant interaction with the In$_2$O$_3$ conduction band minimum. We additionally detect an indirect band gap for In$_2$O$_3$, showing the dispersion of the bulk valence band maximum to be in line with recent theoretical predictions, and observe a reduction in the band gap upon Fe-substitution with possible applications in solar-based photo-devices.

The structural properties and electronic and magnetic characteristics of the particular (In$_{1-x}$Fe$_{x}$)$_{2}$O$_{3}$ samples studied here have been described previously \cite{Xu_Gehring_APL_2009,Xu_Gehring_JAP_2011}. The samples are \mytilde 200 nm thin films prepared by pulsed laser deposition on Al$_{2}$O$_{3}$ (0001) substrates, with Fe content \mbox{$x$ = 0.02,} 0.05, and 0.20. To promote the formation of oxygen vacancies ($V_O$), the samples were synthesized in oxygen deficient environments \cite{SupMat_Green}.  X-ray diffraction studies \cite{Xu_Gehring_APL_2009} showed no secondary metallic or oxide phases and found a decrease in lattice constant from pure In$_{2}$O$_{3}$, consistent with Fe ions substituting for the larger In ions in the lattice. All samples exhibit RTFM with different saturation magnetizations \cite{Xu_Gehring_JAP_2011}.

In Fig. \ref{Fig:EXP}, we depict the unconventional experimental approach used here to obtain the Fe $L_{2,3}$ XAS. Instead of relying on the detection of decay-product electrons or photons from the Fe ions, as is usual in the soft x-ray regime, we realize transmission detection by measuring the x-ray excited optical luminescence (XEOL) intensity originating from the Al$_{2}$O$_{3}$ substrate when incident photons penetrate the film. Such a technique has been used previously with success to measure quantitative XAS signals \cite{Vaz_XEOL_APL_2012, Jakob_PRB_XEOL_2007, Meinert_JPD_XEOL_2011}. Here, using an optical spectrometer, the strong Al$_{2}$O$_{3}$ luminescence line at 689 nm was selected to ensure no signal distortion due to possible weak luminescence effects from the films themselves. In addition to a constant sensitivity throughout the entire film depth, a key feature of such transmission-detected spectra is the elimination of saturation, self-absorption, and non-linear decay effects that usually render secondary yield techniques non-quantitative.

Figure~\ref{In2O3Fe_Iron}(a) displays the raw Fe $L_{2,3}$ XAS spectra of the three (In$_{1-x}$Fe$_x$)$_2$O$_3$ samples. The spectra, which probe primarily the Fe $3d$ electronic properties through $2p$ to $3d$ excitation, are plotted as attenuation coefficients normalized to the In $M_{2}$ edge (and therefore to the In concentration). The XAS spectrum of a powder In$_{2}$O$_{3}$ reference sample with no Fe is also displayed and shows the slightly varying In $M_{2}$-edge background used for normalization between the data sets. The spectra of the Fe-substituted samples exhibit a varying ratio of Fe with valences of 2+ and 3+, most prominently shown through the variation in relative intensity of the two peaks in the $L_{3}$ edge region around \mytilde 709 eV, but also through a general shift of spectral weight to higher energies for higher valence \cite{Crocombette_PRB_1995}. Qualitatively, the lowest concentration sample is found to contain the largest relative amount of Fe$^{3+}$, while there is an increased Fe$^{2+}$ presence for increased Fe concentration. Again, since the spectra were recorded in transmission mode they are not influenced by saturation or other distortion effects and they are thus pure linear combinations of the Fe$^{2+}$ and Fe$^{3+}$ component spectra. 

The Fe component spectra, extracted via spectral differences from the raw data \cite{SupMat_Green}, are shown in Fig. \ref{In2O3Fe_Iron}(b). The Fe$^{2+}$ component can be immediately identified as a fingerprint of octahedrally coordinated, high-spin Fe$^{2+}$, similar (but not identical) to that found in FeO \cite{ReganStohr_PRB_2001, Haup_Haverk_PRB_2010}. The Fe$^{3+}$ component spectrum, however, is very unique, and unlike that found in typical oxides such as Fe$_{2}$O$_{3}$ or Fe$_{3}$O$_{4}$ \cite{ReganStohr_PRB_2001, Crocombette_PRB_1995}. 

To gain insight into the local environment around the Fe ions, we performed full-multiplet single impurity Anderson model (SIAM) calculations for different possible bonding situations of the Fe ion the host \cite{degroot2008}. For each Fe valence, we considered the case of pure $O_h$ symmetry substitution into In sites as well as that of distorted $O_h$ site caused by a neighboring $V_O$ (see supplemental material for details on the SIAM method used \cite{SupMat_Green}). The resulting calculated XAS spectra that best agree with experiment are shown in the lower panel of Fig. \ref{In2O3Fe_Iron}(b).

As one would expect (considering the spectral similarity to FeO), we find that the Fe$^{2+}$ experimental component spectrum is in very good agreement with a SIAM calculation assuming the local $O_h$ symmetry of a substitutional Fe at an In site. Here a small Jahn-Teller distortion is also included to break the orbital degeneracy \cite{SupMat_Green}, as this is typical for the $O_h$, high-spin, $3d^6$ configuration and is similar to what was quantified for Fe$^{2+}$ impurities in MgO \cite{Haup_Haverk_PRB_2010}. For the Fe$^{3+}$ component spectrum, excellent agreement is obtained for the case of Fe substitution at an In site with distortion due to a neighboring $V_O$. Substitution of Fe$^{3+}$ without a neighboring $V_O$ yields a spectrum similar to that of Fe$_2$O$_3$ \cite{SupMat_Green}, which is far from what is observed experimentally as noted above.

To further test these results for the local coordination of the Fe ions, we performed $L_3$ edge RIXS experiments on the samples. Such measurements are similar to XAS in that they probe core-hole mediated excitations, but they give access to a different set of excitations and provide a secondary test of the spectral interpretation. The experimental RIXS results for incident photon energy tuned to the XAS peak at 709.7 eV are shown in Fig. \ref{In2O3Fe_Iron}(c), along with results of SIAM calculations using the same model parameters as for XAS. Here we do not extract component spectra, because varying degrees of self-absorption for different Fe concentrations destroy the pure linear combination of Fe spectra in the case of RIXS. Nonetheless, the agreement between the calculations and experiment is excellent, showing the \mbox{$x = 0.02$} sample contains primarily Fe$^{3+}$-V$_{O}$ with higher energy $dd$ excitations, whereas the \mbox{$x = 0.20$} sample contains primarily Fe$^{2+}$. Thus the identification of each Fe site is well supported by multiple experiments.

Due to the quantitative nature of our XAS measurements, the weight of the component spectra contributing to each raw spectrum of Fig. \ref{In2O3Fe_Iron}(a) can provide the relative concentrations of the Fe$^{2+}$ and Fe$^{3+}$ species for each sample. The extracted concentrations of Fe ions with 2+ and 3+ valences, $x_v$, are plotted against the total Fe concentration $x$ in Fig. \ref{In2O3Fe_Iron}(d). As noted previously, iron is primarily Fe$^{3+}$ in low concentration samples, with increased Fe content yielding more Fe$^{2+}$ sites. On the same plot we show the measured saturation magnetization of each sample ($M_{sat}$, reported previously \cite{Xu_Gehring_JAP_2011}), plotted as $\mu_B$ per (In$_{1-x}$Fe$_x$)$_2$O$_3$ formula unit (F.U.) rather than per Fe ion, so that it may be compared to the amounts of Fe$^{2+}$ and Fe$^{3+}$ per formula unit for each sample on the same plot. Such a comparison reveals a direct correlation between the quantity of Fe$^{3+}$-V$_{O}$ centers and the saturation magnetization, whereas the Fe$^{2+}$ has a very different concentration dependence and does not appear to contribute to the magnetism. The site-specific nature of the XAS has thus allowed us to directly show that the subset of Fe atoms which are adjacent to $V_O$ is responsible for the RTFM. Note that the measured magnetization yields a moment of just under 1 $\mu_{B}/\text{Fe}^{3+}$.

\begin{figure}
\includegraphics[width=3.375in]{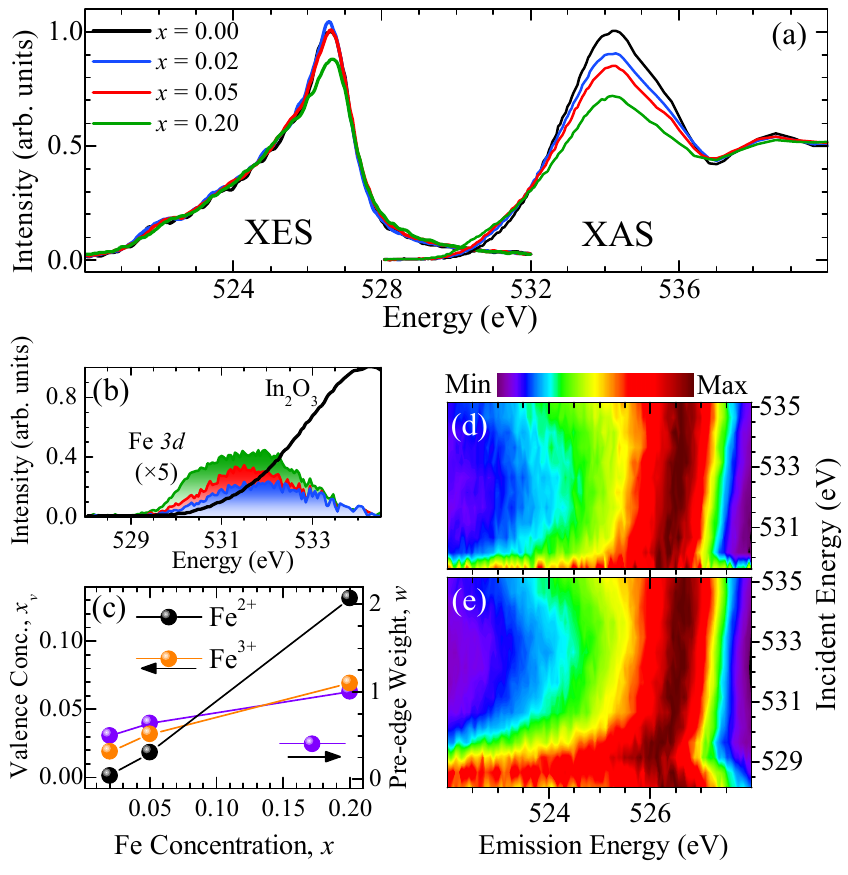}%
\caption{(Color online) Oxygen $K$-edge spectra. (a) XAS and nonresonant XES show the O $2p$ projected CB and VB. (b) The Fe $3d$-derived states near the conduction band minimum are isolated from the XAS. (c) The weight of these Fe $3d$ states correlates directly with the concentration of Fe$^{3+}$. (d) Normalized RIXS spectra with incident energies spanning the absorption pre-edge and onset for the pure In$_2$O$_3$ sample. (e) Corresponding RIXS data for the 5\% substituted sample.}
\label{In2O3Fe_OX}
\end{figure}

Figure \ref{In2O3Fe_OX}(a) displays oxygen $K$-edge XAS and XES of the films, providing respectively the unoccupied and occupied O $2p$--projected DOS on the same energy scale (shifted overall by the $1s$ binding energy). Here we note that although the XAS measures the O $2p$ unoccupied DOS, states of Fe $3d$ character can be probed due to hybridization with the O $2p$ states. Consequently, the XAS reveals in this case the introduction of Fe $3d$ states at the bottom of the conduction band near \mytilde 530 eV for the Fe-substituted samples. These Fe $3d$ electron addition states are isolated from the host In$_2$O$_3$ states by spectral differences, and are displayed for all samples in Fig. \ref{In2O3Fe_OX}(b), along with the spectrum of pure In$_2$O$_3$. Evident from this plot is that the Fe $3d$ states are resonant with the conduction band minimum of In$_2$O$_3$, and their onset energy implies a reduction of the In$_2$O$_3$ band gap by slightly more than 1 eV. 

With the O $K$ XAS normalized far above the edge to the total oxygen content, the isolated Fe states of Fig. \ref{In2O3Fe_OX}(b) provide a quantitative estimate of the near-gap Fe $3d$ electron addition states. The spectral weights of these states for all samples (normalized to the weight for the $x=0.20$ sample) are plotted against the total Fe content in Fig. \ref{In2O3Fe_OX}(c). These pre-edge weights $w$ can then be compared to the quantities of Fe$^{2+}$ and Fe$^{3+}$ determined previously. Like the magnetization in Fig. \ref{In2O3Fe_Iron}(d), we find that the weight of the states resonant with the conduction band minimum correlates very well with the quantity of Fe$^{3+}$ sites. Thus, not only have we found that the Fe$^{3+}$ ions adjacent to oxygen vacancies are responsible for the magnetism, but now we find that the electronic states from these sites are resonant with the conduction band minimum (and Fermi level), providing further support for and insight into the origin of the magnetism. 

To investigate the near-gap region of the DOS in more detail, RIXS spectra were recorded with incident photon energies tuned over the pre-edge region of the XAS. The subsequent RIXS spectral maps for the pure and 5\% substituted samples are shown in Figs. \ref{In2O3Fe_OX}(d) and (e), respectively, with the individual spectra at each incident energy normalized for clarity. First, considering the pure In$_2$O$_3$ sample, we see the RIXS spectra lose all intensity below \mytilde 530 eV incident energy (remembering that the individual spectra are normalized, the thin horizontal band arises due to the strong increase in statistical noise as the RIXS signal drops to zero). This is a consequence of the absence of states in the gap region for this pure sample. We note also the maximum emission energy of the RIXS spectra decreases slightly as the incident photon energy is decreased. This is a signature of the indirect nature of the In$_{2}$O$_{3}$ band gap \cite{Eisebitt_JELSPEC_2000}, and the RIXS map accordingly verifies the presence of a small valence band (VB) dispersion predicted by recent theoretical calculations \cite{Walsh_PRL_2008, King_PRB_2009}. We find an experimental VB dispersion of \mytilde 300 meV whereas the calculations find 50 meV.

For the RIXS map of the 5\% Fe-substituted sample shown in Fig. \ref{In2O3Fe_OX}(e), we see significant changes compared to the pure sample. While again the dispersion of the VB maximum is evident, the RIXS intensity no longer drops to zero below 530 eV incident energy, as there are now Fe $3d$-hybridized states to resonantly excite at this energy. Instead the energy dispersion of the spectrum reverses and just below 528 eV incident energy where the signal now drops to zero, the spectral weight has almost entirely returned to the valence band maximum. This behavior is a clear indication of heavier, less dispersive states at the conduction band minimum, which cause the RIXS to lose $k$-selectivity \cite{SupMat_Green}. This further verifies that the states identified in the pre-edge region are indeed the Fe$^{3+}$-V$_{O}$ impurity band states.

Finally, we note two interesting aspects of the (In$_{1-x}$Fe$_x$)$_2$O$_3$ electronic structure in general. First, as mentioned above, the Fe states lead to a strong reduction of the In$_2$O$_3$ bulk band gap, down toward the high irradiance region of the solar spectrum. This suggests that such materials could be intriguing candidates for solar-based photo-active devices. A second interesting characteristic is the apparent heterovalent substitution at high Fe concentrations--while the magnetically active Fe$^{3+}$-$V_O$ configurations dominate at low concentrations, a saturation occurs and additional Fe has a 2+ valence, substituting for In$^{3+}$. It has indeed been shown theoretically that the 2+ and 3+ charge states of Fe impurities in (fully oxygenated) In$_2$O$_3$ are somewhat close in energy compared to the rest of the $3d$ series \cite{Zunger_TM-In2O3_PRB_2009}. Thus, it is not unreasonable that at the high doping levels the electronic structure is modified enough to favor Fe$^{2+}$. 

In summary, we have shown direct experimental evidence of Fe impurities adjacent to oxygen vacancy centers in ferromagnetic films of Fe-substituted In$_{2}$O$_{3}$. These centers were found to be directly responsible for the room temperature ferromagnetism by using quantitative x-ray absorption spectroscopy. The interaction of the Fe $3d$ states and the host band structure was revealed using oxygen $K$-edge resonant spectroscopy. Our results strongly support the view that a proximity enhanced coupling of Fe dopants to carriers in the impurity band leads to the magnetic behavior.

\begin{acknowledgments}
This work was supported by the Natural Sciences and Engineering Research Council of Canada (NSERC), the Canada Research Chairs program, and the Russian Science Foundation for Basic Research (Project 14-02-00006). XAS measurements were performed at the Canadian Light Source, and RIXS measurements were performed at the Advanced Light Source, which is supported by the U. S. Department of Energy under Contract No. DE-AC02-05CH11231. 
\end{acknowledgments}

\end{document}